\documentclass[showpacs,preprintnumbers,amsmath,amssymb]{revtex4}

\usepackage{epsfig}
\usepackage{graphicx}
\usepackage{dcolumn}
\usepackage{bm}

\let\jnfont=\rm
\def\NPB#1,{{\jnfont Nucl.\ Phys.\ B }{\bf #1},}
\def\PLB#1,{{\jnfont Phys.\ Lett.\ B }{\bf #1},}
\def\EPJC#1,{{\jnfont Eur.\ Phys.\ Jour.\ C }{\bf #1},}
\def\PRD#1,{{\jnfont Phys.\ Rev.\ D }{\bf #1},}
\def\PRL#1,{{\jnfont Phys.\ Rev.\ Lett.\ }{\bf #1},}
\def\MPLA#1,{{\jnfont Mod.\ Phys.\ Lett.\ A }{\bf #1},}
\def\JPG#1,{{\jnfont J.\ Phys.\ G}{\bf #1},}
\def\CTP#1,{{\jnfont Commun.\ Theor.\ Phys.\ }{\bf #1},}
\def\JHEP#1,{{\jnfont JHEP}{\bf #1},}
\begin{document}

\preprint{hep-ph/0608223}

\title{Probing R-parity Violating Interactions from Top Quark Polarization at LHC}

\author{Peiying Li$^1$, Gongru Lu$^1$, Jin Min Yang$^{2,3}$, Huanjun Zhang$^{1,3}$ \\ ~~ \\}

\affiliation{
$^1$ Department of Physics, Henan Normal University, Xinxiang 453007,  China\\
$^2$ CCAST (World Laboratory), P.O.Box 8730, Beijing 100080, China\\
$^3$ Institute of Theoretical Physics, Academia Sinica, Beijing 100080, China }

\date{\today}

\begin{abstract}
In minimal supersymmetric standard model the R-parity violating interactions
can induce anomalous top pair productions at the LHC through the
$t$-channel process $d_R\bar d_R \to t_L\bar t_L$ by exchanging a slepton
or $u$-channel process $d_R\bar d_R \to t_R\bar t_R$ by exchanging a squark.
Such top pair productions with certain chirality cause top quark polarization in the top
pair events. We found that at the LHC, due to the large statistics,  the statistical
significance of the polarization observable and thus the probing ability for the
corresponding R-parity violating couplings are much higher than at the Tevatron upgrade.
\end{abstract}

\maketitle

\section{Introduction}
It has long been speculated that as the heaviest fermion in the Standard Model (SM), the top quark
may have a close connection to new physics \cite{review}. So far there remain plenty
of room for new physics in top quark sector due to the small statistics of the events
discovering the top quark at the Fermilab Tevatron collider.  Since the LHC will produce
top quarks copiously and allow a precision test of top quark nature,  it will
either uncover or stringently constrain the new physics related to the top
quark \cite{sensitive}.

As a popular candidate for new physics, the TeV-scale supersymmetry can sizably
alter some of the top quark properties.  For example, the minimal supersymmetric standard model
(MSSM) can significantly enhance the top quark flavor-changing neutral-current (FCNC) interactions
\cite{t-fcnc} and thus make the top quark FCNC processes possiblly observable at the
LHC.

It is well known that in the MSSM, a discrete multiplicative symmetry of
$R$-parity, defined by $R=(-1)^{2S+3B+L}$ with spin $S$, baryon number $B$
and lepton number $L$, is often imposed on the Lagrangian to maintain the
separate conservation of $B$ and $L$. Since such a conservation requirement
is not dictated by any fundamental principle such as gauge invariance or renormalizability,
the phenomenology of $R$-parity violation has attracted much attention \cite{rv-review}.
So far as the top quark physics at the LHC,  $R$-parity violation can cause some
exotic top quark processes like the $s$-channel single top production and the decays
via exchanging a squark or slepton  \cite{rv-t-prod,rv-t-decay}

Note that the R-parity violating interactions can also induce new mechanisms for top pair
productions
at the LHC through the $t$-channel process $d_R\bar d_R \to t_L\bar t_L$ by exchanging a slepton
or $u$-channel process $d_R\bar d_R \to t_R\bar t_R$ by exchanging a squark.
Although their contribution to the {\it total\/} $t\bar t$ cross section is
unobservably
as small as a few percent \cite{rv-tt}, they may induce a sizable asymmetry between the
left- and right-handed polarized top quarks due to the chiral nature of these couplings.
In \cite{hikasa} such induced polarization in the top pair events is studied
for the Tevatron collider and turns out to be a sensitive probe for
these couplings due to the fact that both the SM and $R$-conserving
MSSM contributions to the polarization are unobservably small \cite{pari-tt}.
Since the upcoming LHC will overwhelmingly outclass the Tevatron
for the study of top quark physics, in this work we extend the analysis of
\cite{hikasa} to the LHC.

This work is organized as follows. In Sec. \ref{sec-2} we briefly describe our calculations.
In Sec. \ref{sec-3} we present some numerical results and give some discussions. Finally,
in Sec. \ref{sec-4} we give our conclusions.

\section{Calculations} \label{sec-2}
The $R$-violating interactions in the superpotential of the MSSM are given by
\begin{eqnarray}\label{WR}
{\cal W}_{\not \! R}
=\frac{1}{2}\lambda_{ijk} L_i L_j E_k^c
+\lambda_{ijk}' \delta^{\alpha\beta} L_i Q_{j\alpha} D_{k\beta}^c
+\frac{1}{2} \lambda_{ijk}'' \epsilon^{\alpha\beta\gamma}
U_{i\alpha}^c D_{j\beta}^c D_{k\gamma}^c +\mu_i L_i H_2 ,
\end{eqnarray}
where $H_{1,2}$ are the Higgs chiral superfields, and $L_i$ ($Q_i$)
and $E_i$ ($U_i$, $D_i$) are the left-handed lepton (quark) doublet
and singlet chiral superfields.
The indices $i$, $j$, $k$ denote generations and
$\alpha$, $\beta$ and $\gamma$ are the color indices.
For top pair productions at hadron colliders,
the lepton number violating couplings $\lambda'_{i31}$ can induce the
$t$-channel process $d_R\bar d_R \to t_L\bar t_L$ by exchanging a slepton
while the baryon number violating couplings $\lambda''_{31j}$ can induce the
$u$-channel process $d_R\bar d_R \to t_R\bar t_R$ by exchanging a squark,
as shown in Fig. 1.
The forms of these amplitudes are discussed in some detail in the
appendix of \cite{hikasa}.
\begin{figure}[hbt]
\epsfig{file=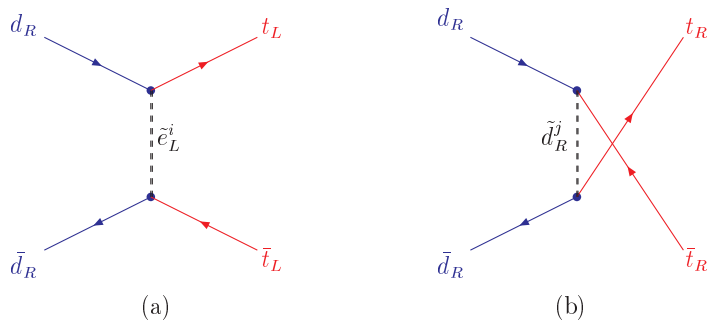,width=10cm}
\caption{ Feynman diagrams for top pair productions
          induced by  $\lambda'_{i31}$ and $\lambda''_{31j}$, respectively.}
\end{figure}
It is clear that due to the chiral nature of these couplings, the processes
they induced in Fig.1 will cause top quark polarization in the $t\bar t$ events,
i.e., an asymmetry between $N_+$ (the number of
$t\bar t$ events with positive helicity of $t$) and $N_-$ (the number of
$t\bar t$ events with negative helicity of $t$). The observables we will
examine are three polarizations: $P_t$ for all $t\bar t$ events and
$P_t^F$ ($P_t^B$) for the $t\bar t$ events with $t$ in the forward (backward)
hemisphere in the $t\bar t$ c.m. frame, defined by \cite{hikasa}
\begin{eqnarray} \label{pt}
P_t  & =&\frac{N_+ - N_-}{N_+ + N_-}=\frac{\sigma_+ - \sigma_-}{\sigma_+ + \sigma_-}, \\
P_t^F& =&\frac{N_+ - N_-}{N_+ + N_-}\left\vert_{\cos\theta^*>0} \right.
       =\frac{\sigma_+ - \sigma_-}{\sigma_+ + \sigma_-}\left\vert_{\cos\theta^*>0} \right. ,\\
P_t^B& =&\frac{N_+ - N_-}{N_+ + N_-}\left\vert_{\cos\theta^*<0} \right.
       =\frac{\sigma_+ - \sigma_-}{\sigma_+ + \sigma_-}\left\vert_{\cos\theta^*<0} \right.,
\label{ptb}
\end{eqnarray}
Here $\theta^*$ is the top scattering angle in the $t\bar t$ c.m. frame,
and $\sigma_+$ ($\sigma_-$) is $t\bar t$ hadronic cross section with positive
(negative) helicity of $t$, which is obtained by convoluting the parton cross
section $\hat\sigma_+$ ($\hat\sigma_-$) with the parton distribution functions
~\cite{cteq} (in our calculations  we assume both the renormalization and factorization
scales to be $\mu=m_t$).

Note that it may seem naive to define the polarization asymmetry as simply
$(N_+ - N_-)/(N_+ + N_-)$ since it would be sizable even when $N_+$ and
$N_-$ are very small (say below unity). However, since $N_+$ ($N_-$) is
the total number of $t\bar t$ events with positive (negative) helicity
and not the number contributed by new physics alone
(the R-violating contribution will become very small for small R-violating
couplings and heavy sfermions involved),
$N_+$ and $N_-$ are very large numbers at the LHC due to the large
QCD $t\bar t$ production cross section and high luminosity of the LHC.
Of course, as will be discussed when we consider the statistical sensitivity
in next section, $N_+$ and $N_-$ will be suppressed by the detection efficiency.
But even so,  $N_+$ and $N_-$ are very large since the LHC can
produce $t\bar t$ copiously and can serve as a top-quark factory.
Also, as will be discussed at the end of this section, although
 $N_+$ and $N_-$ are very large numbers, the  asymmetry
$(N_+ - N_-)/(N_+ + N_-)$ is very small in the SM because
the SM $t\bar t$ production is dominated by QCD processes which
predict $N_+=N_-$. For this reason, this  asymmetry  will be
a sensitive probe for any new physics which can predict a sizable difference
between  $N_+$ and $N_-$.

\begin{figure}[hbt]
\epsfig{file=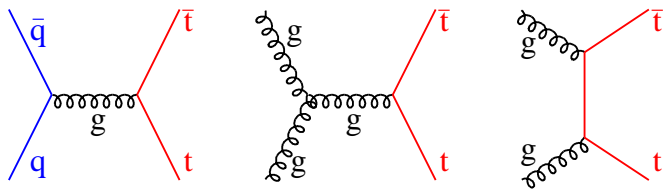,width=10cm}
\caption{ Feynman diagrams for top pair productions in the Standard Model.}
\end{figure}

The parton cross section $\hat\sigma$ contains the SM contribution  $\hat\sigma^{\rm SM}$
and the $R$-violating contribution $\hat\sigma^{\rm new}$
\begin{eqnarray}
\hat\sigma= \hat\sigma^{\rm SM} + \hat\sigma^{\rm new}.
\end{eqnarray}
For $\hat\sigma^{\rm SM}$ the dominant contributions are from the QCD processes shown in
Fig. 2. It is well known that at the LHC (Tevatron) the gluon-gluon fusion process $gg\to t\bar t$
and quark-antiquark annihilation process $q\bar q\to t\bar t$ contribute about $90\%$ ($10\%$) and
$10\%$ ($90\%$), respectively.
We assume that the QCD correction factors (K-factor) to $\hat\sigma^{\rm new}$ and $\hat \sigma^0$
take the same value and thus the QCD correction effects cancel in the polarizations defined
above. Note that due to the separate $C$- and $P$-invariance of QCD, the QCD
cross section  $\hat \sigma^0_+$ equals to  $\hat \sigma^0_-$, which are given by
\begin{eqnarray}
\frac{d\hat\sigma^0_{+}}{d\cos\theta^*}
  =\frac{d\hat\sigma^0_{-}}{d\cos\theta^*}
  = \frac{\pi\alpha_s^2 \beta}{18 \hat s} (1+\cos^2\theta^*)
   +\frac{2\pi\alpha_s^2 \beta m_t^2}{9 \hat s^2} \sin^2\theta^*
\end{eqnarray}
for $q\bar q\to t\bar t$, and
\begin{eqnarray}
\frac{d\hat\sigma^0_{+}}{d\cos\theta^*}
  =\frac{d\hat\sigma^0_{-}}{d\cos\theta^*}
  &=&\frac{\pi\beta\alpha^{2}_{s}}{32\hat{s}}
     \left[12\frac{(\hat{t}-m^{2}_{t})(\hat{u}-m^{2}_{t})}{\hat{s}^{2}}
  -\frac{2}{3}\frac{m^{2}_{t}(\hat{s}-4m^{2}_{t})}
              {(\hat{t}-m^{2}_{t})(\hat{u}-m^{2}_{t})}\nonumber\right.\\
 &&+\frac{8}{3}\frac{(\hat{t}-m^{2}_{t})(\hat{u}-m^{2}_{t})-2m^{2}_{t}(\hat{t}+m^{2}_{t})}
                    {(\hat{t}-m^{2}_{t})^2}
 +\frac{8}{3}\frac{(\hat{t}-m^{2}_{t})(\hat{u}-m^{2}_{t})-2m^{2}_{t}(\hat{u}+m^{2}_{t})}
                  {(\hat{u}-m^{2}_{t})^2}\nonumber\\
 &&\left.+6\frac{(\hat{t}-m^{2}_{t})(\hat{u}-m^{2}_{t})-m^{2}_{t}(\hat{t}-\hat{u})}
          {\hat{s}(\hat{t}-m^{2}_{t})}
 +6\frac{(\hat{t}-m^{2}_{t})(\hat{u}-m^{2}_{t})+m^{2}_{t}(\hat{t}-\hat{u})}
    {\hat{s}(\hat{u}-m^{2}_{t})}\right] ,
\end{eqnarray}
for  $g\bar g\to t\bar t$.
Here $\beta=\sqrt{1-4m^{2}_{t}/\hat{s}}$,
$\hat{t}=m^{2}_{t}-\hat{s}(1-\beta\cos\theta^*)/2$ and
$\hat{u}=m^{2}_{t}-\hat{s}(1+\beta\cos\theta^*)/2$.

The $R$-violating contribution $\hat\sigma^{\rm new}$ comes from the interference of
the diagrams in Fig.1 with the corresponding quark-antiquark annihilation QCD diagram in Fig. 2.
Due to the chiral nature of these  $R$-violating couplings, the cross section with
positive helicity of top quark ($\hat\sigma^{\rm new}_+$) is
not equal to  that with negative helicity ($\hat\sigma^{\rm new}_-$).
The detailed expressions for $d\hat\sigma^{\rm new}_+/d\cos\theta^*$ and
$d\hat\sigma^{\rm new}_-/d\cos\theta^*$ can be found in \cite{hikasa}.

Finally, we stress the reasons to consider the polarization asymmetry
in Eq.(\ref{pt})  rather than
the {\it total\/} $t\bar t$ cross section in probing
the R-violating couplings:
\begin{itemize}
\item[(i)] This polarization asymmetry will be a sensitive probe for new physics
due to the fact that the SM contribution to this polarization asymmetry is very small
(below $1\%$) \cite{pari-tt}.
The $t\bar t$ productions at the LHC are dominated by the QCD processes shown in Fig. 2.
Of course, if we only consider the QCD interactions in the $t\bar t$ productions, the
polarization asymmetry  in Eq.(\ref{pt}) is zero due to the separate $C$- and $P$-invariance
of QCD. The SM electroweak process $q\bar{q} \to Z^* \to t\bar{t}$
can cause a non-zero polarization asymmetry. But the polarization
caused by this process is very small~\cite{pari-tt} since it does not interfere with
the QCD amplitude because of the orthogonal color structure. Furthermore, the polarization
caused by the SM electroweak loop corrections to the  QCD $t\bar t$ production
processes is also very small~\cite{pari-tt}. Therefore, the observation of a sizable
polarization asymmetry (say $P_t >1\%$) will serve as a robust evidence for new
physics.
\item[(ii)] The R-conserving supersymmetric contribution to this polarization asymmetry
is also very small \cite{pari-tt}. So, if supersymmetry is the true story,
only R-violating couplings can cause a sizable value for the polarization asymmetry,
which makes it a unique probe for R-violating couplings.
\item[(iii)] Although the R-violating couplings can
contribute to the {\it total\/} $t\bar t$ cross section by a few percent \cite{rv-tt},
such correction effects of a few percent are hard to be disentangled from
the measurement of the $t\bar t$ cross section
because of the uncertainties in the SM predictions of the $t\bar{t}$
production cross section and the possible loop contributions
from the R-conserving MSSM interactions~\cite{loop}.
\end{itemize}

\section{Results and Discussions} \label{sec-3}
In our numerical calculations the top quark mass $m_{t}=175$ GeV,
and the center-of-mass energy for the LHC (Tevatron) $\sqrt{s}=14$ TeV (2 TeV).
We assume that only one of the involved $R$-violating couplings dominates.
For the $L$-violating couplings $\lambda'_{i31}$, we consider the
existence of $\lambda'_{331}$, in which case the stau is exchanged
in the process, since $\lambda'_{131}$ and
$\lambda'_{231}$ are already strongly constrained by atomic parity
violation and $\nu_{\mu}$ deep-inelastic scattering~\cite{apv}.
For the $B$-violating interactions $\lambda''_{31j}$,
none of them have been well constrained by other processes.
So we may interpret our result as that on any one of $\lambda''_{31j}$,
in which case the squark $\tilde d^j$ is exchanged in the process.

\begin{figure}[hbt]
\epsfig{file=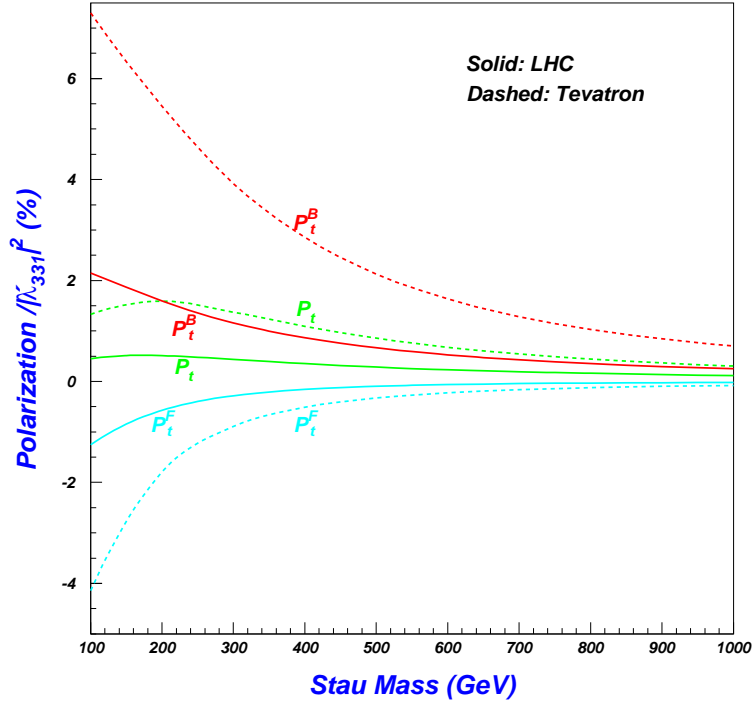,width=10cm}
\caption{ Top quark polarizations $P_t$, $P_t^F$ and  $P_t^B$ normalized
          to $|\lambda'_{331}|^2$ at the LHC ($\sqrt s=14$ TeV)
          and the Tevatron upgrade ($\sqrt s=2$ TeV)
          versus the exchanged stau mass.
      The figure can be also read as  $-P_t$, $-P_t^F$ and  $-P_t^B$
          normalized to $|\lambda''_{31j}|^2$ versus the exchanged squark mass
          with $P_t^F$ and  $P_t^B$ interchanged. }
\end{figure}
In Fig.~3 we show the polarizations $P_t$, $P_t^F$ and $P_t^B$
normalized to $|\lambda'_{331}|^2$ versus the stau mass.
This figure can be also read as  $-P_t$,
 $-P_t^F$ and  $-P_t^B$ normalized to $|\lambda''_{31j}|^2$
versus the mass of the squark $\tilde d^j$ with $P_t^F$ and $P_t^B$
interchanged.
The total polarization is smaller than hemisphere polarizations because
of the cancellation between the two hemispheres.
For the coupling $\lambda'_{331}$ ($\lambda''_{31j}$),
the backward (forward) polarization has the largest magnitude of the three.
Comparing the LHC results with the Tevatron results, we see that the polarizations
at the LHC are smaller. This can be understood as follows:
Compared with the Tevatron, at the LHC both the difference
$\sigma_+ - \sigma_-$ (= $\sigma^{\rm new}_+ - \sigma^{\rm new}_-$) and
the total cross section  $\sigma_+ + \sigma_-$ increase, but the
total cross section increase more significantly since it is dominated
by gluon-gluon fusion process.

Now we estimate the statistical sensitivity.
The statistical error for the top polarization asymmetry defined in Eq.(\ref{pt})
is given by
\begin{eqnarray}
\delta P_t=\frac{[2(N_+^2 + N_-^2)]^{1/2}}{(N_+ + N_-)^{3/2}}
               \simeq \frac{1}{\sqrt{N_+ + N_-}}=\frac{1}{\sqrt{N}}
\end{eqnarray}
while $\delta P_t^F$ ($\delta P_t^B$) is obtained by replacing the
above events numbers with those in forward (backward) hemisphere.
For the $t\bar t$ event number in estimating the statistical error
we take the assumption
\begin{eqnarray}
N= \epsilon \left[ N^{\rm SM}+N^{\rm new} \right]
 \simeq  \epsilon  N^{\rm SM}
 =\epsilon {\cal L} \sigma^{\rm SM}_{t\bar t}
\end{eqnarray}
with $\sigma^{\rm SM}_{t\bar t}
=873$ pb (6.77 pb) for the LHC (Tevatron upgrade) \cite{tt-rate},
${\cal L}$ being the intergrated luminosity
and $\epsilon$ being the efficiency factor in events counting.
The efficiency factor $\epsilon$ can be writen as
\begin{eqnarray}
\epsilon=\epsilon_1\epsilon_2\epsilon_3
\end{eqnarray}
with $\epsilon_{1,2,3}$ coming respectively from three aspects:
\begin{itemize}
\item[(1)] Firstly, as discussed in many previous works \cite{pola1,pola2,pola3},
the helicity states of the top quark may have to be determined from the kinematic
distribution of its decay particles $t \to W^+ b \to \ell^+\nu_\ell b$
($\ell=e$ or $\mu$) after reconstructing the top and anti-top quarks in the
signal events. This will immediately lead to a suppression factor
$\epsilon_1=2/9$.

\item[(2)] Secondly, as is well known,
considering both the detector acceptance and the suppression of
the QCD background, some kinematic cuts on the top and anti-top decay
particles as well as $b$-tagging are necessary.
Depending on the detailed kinematic cuts and the $b$-tagging efficiency,
the events number is further suppressed by an
efficiency factor $\epsilon_2$.

For example, if we impose the kinematic selection cuts as
$p_T^{\ell}, p_T^{jet} \ge 20$ GeV, $ \eta_{jet},~\eta_{\ell} \le 3.0$,
$p_T^{\rm miss}\ge 30$ GeV and $\Delta R_{jj},~\Delta R_{j\ell} \ge 0.4$,
where $p_{T}$ denotes the transverse momentum, $\eta$ is the pseudo-rapidity,
and $\Delta R$ is the separation in the azimuthal angle-pseudo rapidity plane
$(~\Delta R= \sqrt{(\Delta \phi)^2 + (\Delta \eta)^2}~ )$ between
a jet and a lepton or between two jets, then the efficiency from such
kinematic cuts is about $30\%$ \cite{stop}.  Combined with the b-tagging efficiency
of about $50\%$, the efficiency factor $\epsilon_2 \simeq 15\%$.
(Currently at Tevatron CDF,  the efficiency for tagging at least one b-jet in a $t\bar t$ event
is $(53\pm 4)\%$ \cite{tt-CDF}. At the LHC we expect an efficiency better than
$50\%$ for tagging at least one b-jet in a $t\bar t$ event.)

Note that in theoretical analyses we usually impose $\Delta
R_{jj},~\Delta R_{j\ell} \ge 0.4$ \cite{stop}. But for the
separation of jets at the LHC, $\Delta R_{jj} \ge 0.4$ may be
over-optimistic and the realistic cut may be $\Delta R_{jj} \ge
0.7$. As a result of such a stronger cut, the efficiency factor
$\epsilon_2 $ may drop below $10\%$.

Further, in theoretical analyses we usually imposed $\eta\le 3$ \cite{stop}
and did not consider the difference of detector efficiency in the different
regions of  $\eta\le 3$.
In practice, the detector efficiency will certainly be quite different in
    the barrel $\eta < 1$ and endcap $1 < \eta < 3$ regions.
Therefore, the realistic value of $\epsilon_2$ can only be obtained
from detector-dependent full Monte Carlo simulation.

\item[(3)] Thirdly, from analysing the distribution of the leptonic decays of a
top quark, we in practice can only know its helicity with some probability
$\epsilon_3$. A recent analysis \cite{pola2} gives $\epsilon_3 \simeq 75\%$.
But this figure may be over-optimistic since the analysis  \cite{pola2} is
a theoretical work making rather optimistic assumptions.
As pointed in  \cite{pola3},
the principal difficulty in measuring top quark
polarization comes from the ambiguities in reconstructing its momentum
from the decay products, when both top and anti-top decay
semileptonically. The realistic value of $\epsilon_3$ may also have to be
obtained from detector-dependent full Monte Carlo simulation.
For the ATLAS detector an analysis was given in \cite{atlas} based on
leading-order Monte Carlo generators and on a fast simulation of the detector.
\end{itemize}

Since $\epsilon_{2,3}$ are detector-dependent and their realistic
values can only be obtained from detector-dependent full Monte Carlo simulations,
in our numerical calculations for statistical significance and statistical limits
we only incorporate the suppression factor $\epsilon_1$ while leave $\epsilon_{2,3}$
as variables. Of course, it is straitforward to reinterpret our results once
$\epsilon_{2,3}$ are determined. We will explicitly point out how to
reinterpret our results once $\epsilon_{2,3}$ are known.

\begin{figure}[hbt]
\epsfig{file=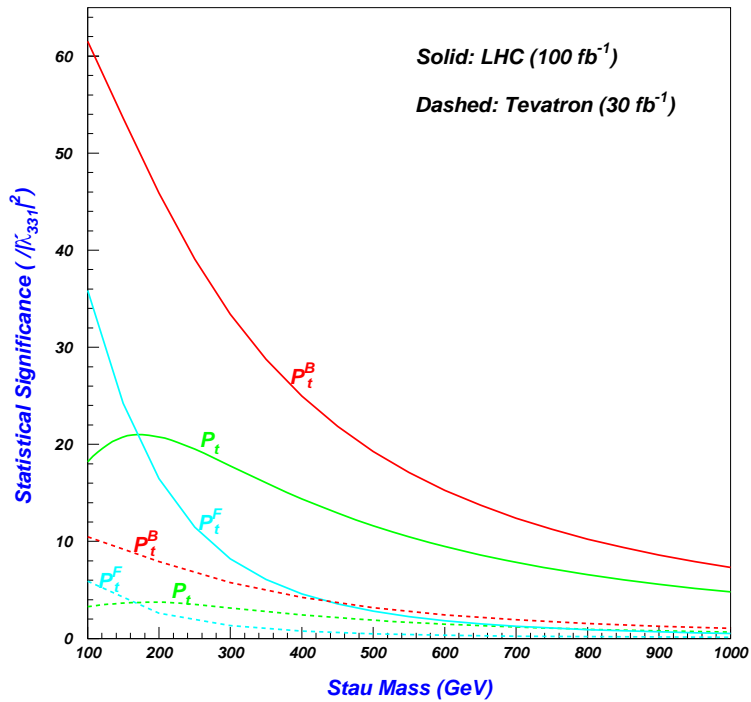,width=10cm}
\caption{ Statistical significance at the LHC ($\sqrt s=14$ TeV)
          and the Tevatron upgrade ($\sqrt s=2$ TeV),  normalized
          to $|\lambda'_{331}|^2$
          (or to $|\lambda'_{331}|^2 \sqrt{\epsilon_2 \epsilon_3}$ if
           we consider the efficiency factors $\epsilon_{2,3}$).
          The figure can be also read as the statistical significance
          normalized to $|\lambda''_{31j}|^2$
      versus the exchanged squark mass with $P_t^F$ and  $P_t^B$ interchanged.}
\end{figure}
The statistical significances $P_t/\delta P_t$, $P_t^F/\delta P_t^F$ and
$P_t^B/\delta P_t^B$ normalized to $|\lambda'_{331}|^2$ versus stau mass
are shown in Fig. 4.
The figure can be also read as the statistical significances
normalized to $|\lambda''_{31j}|^2$ versus the exchanged squark mass
with $P_t^F$ and $P_t^B$ interchanged.
We see that in case of $\lambda'_{331}$ ($\lambda''_{31j}$)
the sensitivity of the backward (forward) polarization is the best.
Comparing Fig. 3 with Fig. 4, we see that although the polarizations at the LHC
are smaller than at the Tevatron upgrade, the LHC gives much larger statistical
significances due to its much larger $t\bar t$ sample.

The statistical limits for $\lambda'_{331}$
($\lambda''_{31j}$) versus stau (squark) mass are shown in Fig.~5.
Here for the LHC we show the limits at $5\sigma$ (discovery), $3\sigma$ (evidence)
and  $2\sigma$ (exclusion).
The current $2\sigma$ exclusion limits from $Z$ decay at LEP~\cite{Zdecay}
are also shown for comparison. We see that the LHC may be quite powerful in
probing these $R$-violating couplings.

\begin{figure}[hbt]
\epsfig{file=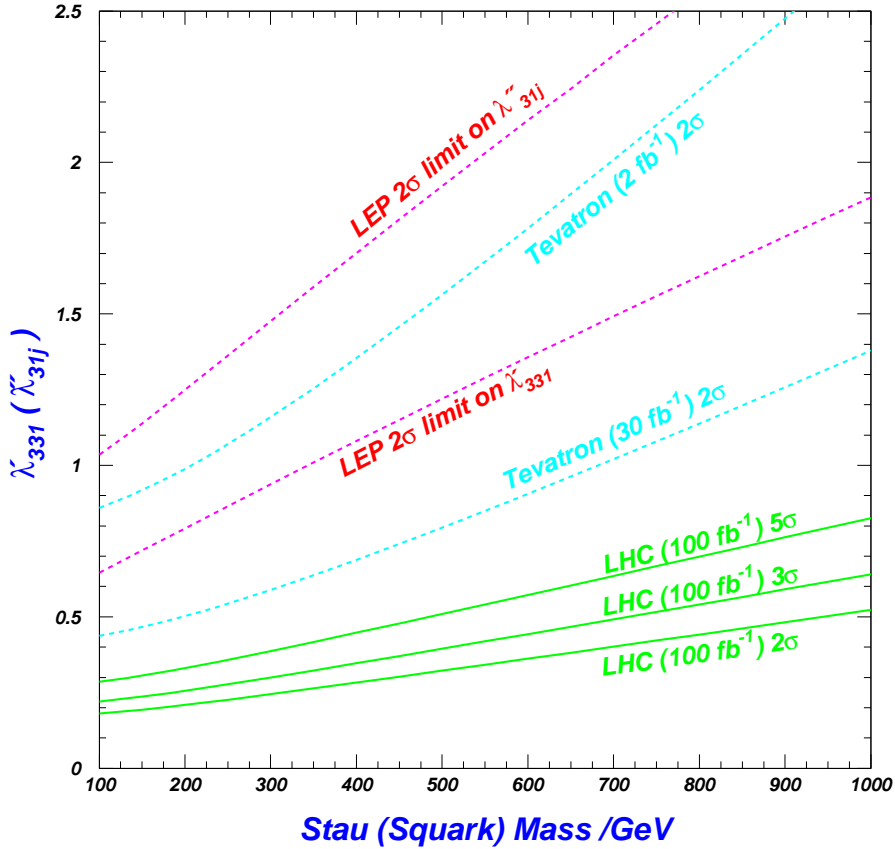,width=12cm,angle=0}
\caption{ The statistical limits for $\lambda'_{331}$
          (or for $\lambda'_{331} (\epsilon_2 \epsilon_3)^{1/4}$
           if we consider the efficiency factors $\epsilon_{2,3}$)
           versus stau mass.
           The figure can be also read as the statistical limits for
           $\lambda''_{31j}$ versus squark mass.}
\end{figure}

Finally, we stress that our above analyses are almost solely
theoretical. We only demonstrated the statistical sensitivity and
statistical limits. As is well known, in experiments some
systematic errors are inevitable. Such systematic errors are
detector-dependent and will certainly worsen the probing limits
for the R-violating couplings. For example, as an artifact of the
measurement problems, an apparent polarization asymmetry may
arise. The analysis for the ATLAS detector showed \cite{atlas}
that the systematic uncertainties will dominate over the
statistical errors after one LHC year at low luminosity (10
fb$^{-1}$). The analysis in \cite{atlas} gives a quite encouraging
result, which shows that a percent-level precision on the
measurement of the polarization asymmetry between like-spin versus
unlike-spin $t\bar t$ events is possible even with 10 fb$^{-1}$.
Of course, a higher luminosity (say 100 fb$^{-1}$) will further
enhance the measurement precision. Although the analysis in
\cite{atlas} did not show the possible measurement precision on
our polarization asymmetry $(N_+ - N_-)/(N_+ + N_-)$, we can
expect a similar or even better precision because here we sum over
the helicities of the $\bar{t}$ and consider an integrated
asymmetry in the numbers of $t_+$ and $t_-$ produced.

\section{Conclusions} \label{sec-4}
In summary, the R-parity violating interactions of the top quark, which have not been well
constrained by current experiments, can induce anomalous top pair productions
at hadron colliders. Although such induced processes only contribute to the
total $t\bar t$ cross section at the percent level, they can cause top quark polarization
in the top pair events due to the chiral nature of these interactions.
The polarization can be a useful observable for probing these interactions at
the LHC and the upgraded Fermilab Tevatron collider because the polarization is expected
to be very small in the Standard Model.
We found that at the LHC, due to the large statistics,  the statistical
significance of the polarization observable and thus the probing ability for the
corresponding R-parity violating couplings are much higher than at the Tevatron
upgrade.
Finally, we stress that for the purpose of probing these R-vioalting couplings,
there may exist other processes which are complementary to our analyses.
For example, $\lambda''_{3ij}$ couplings can induce $s$-channel top-squark
productions at the LHC and Tevatron
and, as studied in \cite{one-stop}, such productions can be used to
probe the parameter space of $\lambda''_{3ij}$ versus top-squark mass.


\begin{thebibliography}{99}

\bibitem{review} For some reviews on top quark physics, see, e.g.,
               D. Chakraborty, J. Konigsberg, D. Rainwater,  Ann. Rev. Nucl. Part. Sci.
                                                             {\bf 53}, 301  (2003);
               E.~H.~Simmons, hep-ph/0211335;
               C.-P. Yuan,  hep-ph/0203088;
               S. Willenbrock, hep-ph/0211067.
               M. Beneke {\it et al.}, hep-ph/0003033.
\bibitem{sensitive}  For model-independent studies of new physics in top quark physics,
               see, e.g.,
               C. T. Hill and S. J. Parke, \PRD49, 4454 (1994);
               K. Whisnant, et al.,  \PRD56, 467 (1997);
               K. Hikasa, et al., \PRD58, 114003 (1998).
\bibitem{t-fcnc}    C.~S.~Li, R.~J.~Oakes and J.~M.~Yang, \PRD49, 293 (1994);
                    G.~Couture, C.~Hamzaoui and H.~Konig, \PRD52, 1713 (1995);
                    J.~L.~Lopez, D.~V.~Nanopoulos and R.~Rangarajan, \PRD56, 3100  (1997);
                    G.~M.~de Divitiis, R.~Petronzio and L.~Silvestrini, \NPB504, 45 (1997).
                    J.~M.~Yang, B.-L.~Young and X.~Zhang, \PRD58, 055001 (1998);
                    J.~M.~Yang and C.~S.~Li, \PRD49, 3412 (1994);
                    J.~Guasch, and J.~Sola, \NPB562, 3 (1999);
                    G. Eilam, {\it et al.}, \PLB510, 227 (2001);
                    J. Cao, {\it et al.},  \NPB651, 87 (2003);
                    J. J. Liu, C. S. Li, L. L. Yang, L. G. Jin, \PLB599, 92 (2004);
                    J. Cao,  {\it et al.}, \PRD74, 031701 (2006);
                    G. Eilam, M. Frank, I. Turan, hep-ph/0601253.
\bibitem{rv-review} For some reviews, see
                 R. Barbier, et al., Phys.\ Rept.\  {\bf 420}, 1 (2005);
                 H. Dreiner, hep-ph/9707435;
                 G. Bhattacharyya, hep-ph/9709395;
                 P. Roy, hep-ph/9712520.
\bibitem{rv-t-prod} A. Datta,  {\it et al.}, \PRD56, 3107 (1997);
                     R. J. Oakes, {\it et al.}, \PRD57, 534 (1998);
                     P. Chiappetta, {\it et al.}, \PRD61, 115008 (2000).
\bibitem{rv-t-decay} K.J. Abraham, {\it et al.}, \PRD63, 034011 (2001); \PLB514, 72 (2001).
\bibitem{rv-tt}  D.~K. Ghosh, S.~Raychaudhuri and K.~Sridhar, \PLB396, 177 (1997).
\bibitem{hikasa} K. Hikasa, J. M. Yang and B.-L. Young, \PRD60, 114041 (1999).
\bibitem{pari-tt}
      C.~Kao, G.~A. Ladinsky and C.-P. Yuan, FSU-HEP-930508, 1993 (unpublished);
      C.~Kao, \PLB348, 155 (1995);
      C.~S. Li, {\it et. al.}, \PLB398, 298 (1997);
      C.~S. Li, C.-P. Yuan and H.-Y. Zhou, \PLB424, 76 (1998).
\bibitem{cteq} J.~Pumplin, {\it et. al.},  JHEP {\bf 0207}, 012 (2002);
               D.~Stump, {\it et. al.}, JHEP {\bf 0310}, 046 (2003).
\bibitem{loop}   C.~S. Li,  {\it et al.}, \PRD52, 5014 (1995);
J. Kim, J.~L. Lopez, D.~V. Nanopoulos and R.~Rangarajan, \PRD54, 4364 (1996);
J.~M. Yang and C.~S. Li, \PRD52, 1541 (1995); {\bf 54}, 4380 (1996);
C.~S. Li,  {\it et al.}, \PLB379, 135 (1996);
S.~Alam, K.~Hagiwara and S.~Matsumoto, \PRD55, 1307 (1997);
Z.~Sullivan, \PRD56, 451 (1997);
W.~Hollik, W.~M. Mosle and D.~Wackeroth, \NPB516, 29 (1998).

\bibitem{apv}  V.~Barger, G.~F. Giudice and T.~Han, \PRD40, 2978 (1989).
\bibitem{pola1} G.~L. Kane, G.~A. Ladinsky and C.-P. Yuan, \PRD45, 124 (1992);
               M.~Jezabek and J.~H. K\"uhn, \PLB329, 317 (1994);
               S.~Parke and Y.~Shadmi, \PLB387, 199 (1996);
\bibitem{pola2}  Y. Sumino and S. Tsuno, \PLB633, 715 (2006).
\bibitem{pola3}  R. M. Godbole, S. D. Rindani, R. K. Singh,  \JHEP0612, 021 (2006).
\bibitem{atlas}  F. Hubaut, E. Monnier, P. Pralavorio, V. Simak, K. Smolek, hep-ex/0508061.
\bibitem{tt-rate} See, e.g., N. Kidonakis and R. Vogt, \PRD68, 114014 (2003);
                                                        \EPJC33, s466 (2004).
\bibitem{Zdecay}  J.~M. Yang, \EPJC20, 553 (2001);
                  G.~Bhattacharyya, J.~Ellis and K.~Sridhar, \MPLA10, 1583 (1995);
                  G.~Bhattacharyya, D.~Choudhury and K.~Sridhar, \PLB355, 193 (1995).
\bibitem{stop} See, e.g., J. M. Yang, B.-L. Young, \PRD62, 115002 (2000);
                     M. Hosch, et al., \PRD58, 034002 (1998);
                     R. J. Oakes, et al., \PRD57, 534 (1998).
\bibitem{tt-CDF} See, e.g., H. Bachacou, hep-ex/0501007.
\bibitem{one-stop} E. L. Berger, B. W. Harris, Z. Sullivan,
                   \PRL 83, 4472 (1999); \PRD 63, 115001 (2001).

\end{thebibliography}
\end{document}